\documentclass[pre,onecolumn]{revtex4-2}
\usepackage{amsmath}
\usepackage{bm}
\usepackage{dsfont}
\usepackage[left = 1.4cm, right = 1.4cm, top = 1.8cm, bottom = 1.8cm]{geometry}

\usepackage{graphicx}

\usepackage{enumitem}

\usepackage{color}

\definecolor{OliveGreen}{rgb}{0.1, 0.4, 0.1}

\usepackage{setspace}
\emergencystretch=\maxdimen
\newcommand{\ch}{\operatorname{ch}}
\newcommand{\sh}{\operatorname{sh}}

\newcommand{\csh}{\operatorname{csh}} 
\newcommand{\sch}{\operatorname{sch}} 

\newcommand{\sgn}{\operatorname{sgn}}

\newcommand{\evarthe}{\hat{ \bm{e} }_\vartheta}

\newcommand{\evarphi}{\hat{ \bm{e} }_\varphi}

\allowdisplaybreaks

\usepackage[colorlinks=true,citecolor=OliveGreen,linkcolor=OliveGreen,urlcolor=OliveGreen]{hyperref}

\usepackage{mathrsfs}

\begin{document}

\title{Phoretic interactions in two-medium wedge geometries}

\author{Abdallah Daddi-Moussa-Ider}%
\email{abdallah.daddi-moussa-ider@open.ac.uk}
\thanks{Corresponding author}
\affiliation{School of Mathematics and Statistics, The Open University, Walton Hall, Milton Keynes MK7 6AA, United Kingdom}

\begin{abstract}
We investigate the diffusiophoretic motion of a chemically isotropic active colloid in a three-dimensional wedge formed by two distinct fluid media, in the limit of vanishing Péclet and Reynolds numbers. The concentration field is obtained using the Fourier--Kontorovich--Lebedev transform, yielding an exact representation for arbitrary wedge opening angles and interfacial contrasts. We introduce the interfacial parameter $\Gamma=(1-\lambda\ell)/(1+\lambda\ell)$, where~$\lambda$ denotes the diffusivity contrast and~$\ell$ the solute partition coefficient. For $\Gamma=\pm1$ and commensurate wedge angles, the solution reduces to finite image constructions, with distinct structures for even and odd commensurability. The general solution also recovers the planar-interface and semi-infinite-interface limits.
The leading-order translational phoretic velocity is derived from the concentration field, revealing a strong interplay between wedge geometry and interfacial properties that governs both the magnitude and direction of particle motion. This work provides a framework for understanding and controlling phoretic transport in confined multiphase environments and offer a basis for extensions to finite-size geometries and mixed fluid--fluid and solid boundary conditions.
Our results may find applications in the control of active-particle transport in confined multiphase environments, where interfacial properties and geometry can be exploited to tune phoretic motion.
\end{abstract}

\maketitle

\section{Introduction}

Active matter has rapidly expanded into a dynamic, interdisciplinary field spanning biophysics, soft matter, and bioengineering~\cite{Chen2025,lauga09,zottl16,bechinger16,zottl2023modeling,te2025metareview}. At its core, active matter concerns systems composed of individual or collective constituents that consume energy to generate and sustain motion, giving rise to rich non-equilibrium dynamics that have no direct equilibrium counterpart. A central class of these systems comprises self-propelled microswimmers, simplified yet powerful platforms for probing non-equilibrium processes relevant to biological and cellular contexts. By continuously converting ambient energy into mechanical work, microswimmers sustain autonomous locomotion without external driving, enabling them to explore complex fluid environments and interact dynamically with surrounding boundaries, interfaces, and other active or passive objects. Their ability to operate out of equilibrium also provides a framework for investigating transport, collective motion, and emergent phenomena at microscopic scales. Beyond fundamental interest, these capabilities hold strong potential for biomedical and technological applications, including targeted drug delivery, microscale surgery, and diagnostics, where controlled motion at small length scales is essential~\cite{park2017multifunctional,tang2020enzyme,soto2020medical,llacer2021biodegradable}.

Phoretic propulsion offers a general mechanism for autonomous motion in active colloidal systems~\cite{moran17,illien17, golestanian_les_houches, liebchen2022interactions}. Rather than relying on external forces or imposed flow fields, phoretic particles exploit physicochemical processes occurring at their surfaces to modify the local chemical environment. Surface reactions generate spatially varying solute concentrations, which in turn produce gradients along the particle surface and induce phoretic slip flows that propel the particle through the surrounding fluid~\cite{golestanian05,golestanian07,michelin14}. This mechanism is particularly attractive because it provides a direct route from microscopic surface chemistry to macroscopic particle motion, while allowing propulsion to emerge spontaneously from the coupling between chemical transport and hydrodynamics. In realistic environments, however, the surrounding geometry can substantially modify this coupling. Boundaries, interfaces, and confining walls perturb both the chemical and hydrodynamic fields generated by an active particle, thereby altering its translational and rotational dynamics~\cite{uspal15,ibrahim15,ibrahim2016walls,das2015boundaries,simmchen16,uspal16,mozaffari16}. Depending on the confinement geometry, surface properties, and flow conditions, such boundary effects can lead to qualitatively different behaviors, including surface-following trajectories, reorientation, stable hovering near a boundary, and even upstream migration of spherical particles~\cite{sharan2022upstream}. The presence of multiple particles introduces an additional layer of complexity, as each swimmer responds to chemical and hydrodynamic disturbances generated by its neighbors. These mutually mediated fields can produce long-range interactions and nontrivial pair dynamics, ultimately giving rise to collective behaviors and emergent dynamical states that are absent for isolated particles~\cite{sharifi16, saha2019pairing, varma2019modeling,nasouri2020exact,sharan2023pair}.
Collective dynamics of phoretically interacting active particles has been shown to exhibit dynamic clustering and chemotactic collapse through self-generated chemical fields~\cite{pohl2014dynamic}, while chemo-repulsive interactions in the presence of pinning disorder can give rise to flocking transitions and polar order~\cite{adhikary2026flocking}.

Spatial symmetry plays a fundamental role in determining whether a phoretic particle can undergo spontaneous locomotion in low-Reynolds-number flows. In the absence of a preferred direction, a perfectly symmetric particle in an unbounded and homogeneous environment cannot sustain net translation, as the chemical and hydrodynamic fields retain the symmetries of the particle and its surroundings. Importantly, however, this symmetry can be broken not only through chemical activity but also by the particle geometry or its external environment. Thus, even for chemically isotropic surfaces, geometric or environmental asymmetries can distort the surrounding chemical field and generate tangential solute gradients capable of producing a net propulsive motion. For example, isotropic active particles can acquire motility through phoretic and hydrodynamic coupling with passive obstacles or self-organized clusters, where the surrounding configuration provides the asymmetry required for spontaneous propulsion~\cite{soto2014self,michelin2015autophoretic,varma2018clustering,picella2022confined}. Similarly, confinement and nearby boundaries can break the spatial symmetry of the chemical and flow fields, leading to directed motion even when the particle itself is chemically isotropic. Asymmetric particle geometry provides another intrinsic route to symmetry breaking: deviations from spherical or otherwise fore-aft-symmetric shapes can generate nonuniform solute distributions and induce phoretic flows across chemically uniform surfaces~\cite{michelin2015geometric,lisicki2016phoretic,nowak2025diffusioosmotic,daddi2026phoretic}. These examples highlight that spontaneous phoretic locomotion does not necessarily require chemical heterogeneity; rather, it can emerge from the interplay between surface activity, particle geometry, and environmental symmetry.

In this work, we employ a far-field framework to investigate the phoretic interactions generated by a monopolar source in wedge domains comprising two distinct media, bounded by straight, semi-infinite surfaces that meet at a common wedge apex. This geometry provides a natural model for confined environments in which a fluid–fluid interface or a pair of intersecting boundaries introduces a localized geometric singularity, and captures the leading-order diffusiophoretic response of a chemically isotropic, catalytically active spherical colloid placed within such a two-medium wedge. By focusing on the monopolar contribution, we isolate the dominant long-range component of the solute field and thereby obtain analytical expressions that reveal how the wedge geometry and material contrast influence the resulting phoretic interactions. Although diffusiophoretic motion near no-slip boundaries~\cite{uspal15,ibrahim15,ibrahim2016walls,das2015boundaries,simmchen16,uspal16,mozaffari16} and near fluid--fluid interfaces~\cite{malgaretti2018self,daddi2022diffusiophoretic} has received considerable attention, the corresponding behavior in the vicinity of sharp corners and wedge-like confinements remains relatively unexplored. In particular, the combined effects of confinement and the presence of two distinct media can lead to nontrivial modifications of the chemical field and, consequently, of the particle dynamics. The present framework therefore provides a convenient setting for examining how wedge geometry and interfacial properties govern phoretic interactions and particle motion near sharp boundaries.

From a hydrodynamic perspective, the behavior of fields near corners has a long history, motivated by the singular and strongly geometry-dependent flow structures that arise in the vicinity of intersecting boundaries. Early studies examined Stokes flows in wedge configurations, establishing the characteristic scaling and spatial structure of viscous flows near sharp corners~\cite{sano76,sano1977slow,sano1978effect,hasimoto80}. These works demonstrated that the local flow field can differ qualitatively from that near smooth boundaries, with the wedge angle playing a central role in determining the nature of the hydrodynamic response. The leading-order structure of three-dimensional Stokes flow near a corner was subsequently revisited using techniques from complex analysis, providing a deeper understanding of the asymptotic behavior of the velocity and pressure fields and their dependence on the local geometry~\cite{dauparas2018leading,dauparas2018stokes}. More recently, the method of images has been employed to investigate the motion of microswimmers confined by wedge-shaped free-slip interfaces for commensurate wedge opening angles, highlighting the rich dynamical behavior that can emerge from corner-induced hydrodynamic interactions~\cite{sprenger2023microswimming}. Related developments have also led to the construction of Green's functions for homogeneous, isotropic, linearly elastic wedges subject to a variety of boundary conditions~\cite{daddi2025proc,daddi2025jelasticity}. These results are particularly relevant because, in the incompressible limit, the governing equations for linear elasticity acquire the same mathematical structure as those describing low-Reynolds-number viscous flows. Consequently, the corresponding Green's functions exhibit a direct mathematical connection, allowing analytical techniques developed for one class of problems to inform the treatment of the other.

In our recent work~\cite{daddi2026selfdiffusio}, we investigated self-diffusiophoretic propulsion in a wedge confined by two impermeable walls subject to no-flux boundary conditions, with particular emphasis on the role of phoretic interactions in modifying the particle dynamics. Here, we extend this framework to a more general and physically relevant configuration in which the wedge boundaries are fluid--fluid interfaces separating two media with different solute diffusivities. This extension introduces an additional source of asymmetry into the problem, since the solute fields on either side of each interface are coupled through interfacial continuity conditions while evolving with different diffusivities. As a result, the phoretic response becomes sensitive not only to the wedge geometry but also to the diffusivity contrast between the two media, providing a means to control the strength and direction of the resulting particle motion. Wedge-shaped geometries and sharp corners are also ubiquitous in multiphase and biological systems, where interfaces and boundaries frequently intersect over microscopic length scales. For example, microswimmers or active particles confined within droplets can encounter wedge-like regions formed by intersecting fluid--fluid interfaces, particularly in multiphase and emulsion systems. Similar wedge geometries arise near fluid--fluid interfaces in confined microfluidic environments and at junctions between immiscible fluid domains. In these settings, the interplay between the wedge geometry and the contrast in interfacial properties can strongly influence the surrounding concentration and flow fields, thereby providing a relevant framework for studying particle transport and propulsion in multiphase fluids. Characterizing phoretic transport in such configurations is therefore important for understanding how confinement, interfacial properties, and material contrasts collectively regulate the propulsion, migration, accumulation, and trapping of active particles near corners and contact lines.

In this work, we employ the Fourier--Kontorovich--Lebedev (FKL) transform, which provides a natural and powerful framework for solving boundary-value problems in geometries that combine radial structure with angular confinement, such as wedges and cones. The integral transform allows the spatial dependence of the governing equations to be separated in a manner well suited to the intersecting boundaries of the wedge, while providing a convenient representation of the resulting Green's functions and solute fields. Our analysis centres on the fundamental monopolar solution, which represents the leading-order contribution to the solute field generated by a localized source and serves as a basic building block for describing more general surface activity distributions on phoretic particles. By establishing this elementary solution in the two-medium wedge geometry, we obtain an analytically accessible representation from which more complex source distributions and particle configurations can subsequently be constructed. 
Exact analytical expressions are obtained for the limiting values of the interfacial parameter, $\Gamma=\pm1$, as well as for commensurate wedge angles.
We then use the monopolar solution to derive the self-diffusiophoretic velocity of a chemically isotropic active particle and examine how its motion is governed by the geometry and material properties of the surrounding domain. In particular, we focus on the effects of the wedge opening angle and the particle's position within the wedge, which together determine the degree of confinement and the symmetry of the solute field experienced by the particle. This analysis provides a clear understanding of how geometric confinement and diffusivity contrast modify the phoretic response of an otherwise isotropic active particle.

\section{Mathematical model}

\begin{figure}
    \centering
    \includegraphics[width=0.6\linewidth]{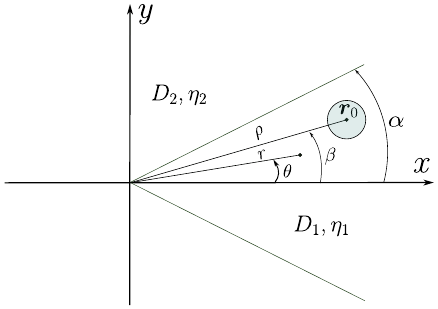}
    \caption{A catalytically isotropic active particle of radius $R$ is situated within a two-medium wedge domain, where the two fluids are separated by a fluid–fluid interface intersecting along the $z$-axis and forming a wedge of semi-opening angle $\alpha \in (0,\pi]$. The evaluation point is specified by the cylindrical coordinates $(r,\theta,z)$, while the position of the active colloid is characterized by its polar distance $\rho$ and polar angle $\beta$.
    The medium in which the active colloid is located is characterized by viscosity $\eta_1$ and solute diffusivity $D_1$, while the adjoining medium is characterized by viscosity $\eta_2$ and solute diffusivity $D_2$.
}
    \label{fig:illus}
\end{figure}

A chemically isotropic, self-propelling spherical colloid of radius $R$ is suspended in a three-dimensional viscous fluid within a wedge geometry comprising two distinct media, as illustrated in Fig.~\ref{fig:illus}. Using cylindrical coordinates $(r,\theta,z)$ with the wedge edge aligned with the $z$-axis, the interfaces are defined by $\theta=\pm\alpha$, with $\alpha\in(0,\pi]$. Fluid 1, of viscosity $\eta_1$, occupies the interior wedge, $|\theta|\leq\alpha$, while fluid 2, with viscosity $\eta_2$, fills the complementary region, $\alpha\leq|\theta|\leq\pi$. The particle is centered at $\bm{r}_0=(\rho,\beta,0)$ within fluid 1.

We work in the low-Reynolds-number regime, where viscous forces dominate over inertia and the surrounding flow is described by Stokes hydrodynamics~\cite{happel12}. We further consider the zero-Péclet-number limit, under which solute transport is purely diffusive and advection by the particle-induced flow can be neglected. These approximations are well suited to the dynamics of microscale active colloids in typical fluid environments, for which both inertial and advective transport effects are generally weak.

\subsection{Governing equations}

In the far-field description, the particle is represented by a point source of strength $4\pi$, and we seek the resulting concentration fields $c_1$ and $c_2$ in the two media. These fields satisfy the corresponding Poisson equation with a singular source at $\bm{r}_0$, together with interfacial conditions at $\theta=\pm\alpha$ enforcing continuity of solute flux and partitioning of the concentration between the two media.
The governing partial differential equations are
\begin{equation}
    \boldsymbol{\nabla}^2 c_1 + \frac{ 4\pi }{r}\, \delta(r-\rho)\,\delta(\theta-\beta)\,\delta(z) = 0 \quad \text{ for }\, |\theta|\le \alpha \, , \qquad
    \boldsymbol{\nabla}^2 c_2 = 0 \quad \text{ for }\, \alpha\le|\theta|\le\pi \, . 
\end{equation}
These equations are subject to the boundary conditions
\begin{equation}
    \ell c_1 = c_2 \, , \qquad
    D_1 \, \frac{\partial c_1}{\partial \theta} = 
    D_2 \, \frac{\partial c_2}{\partial \theta} 
 \qquad \text{ at } \theta =\pm\alpha \, ,
    \label{eq:BCs}
\end{equation}
where $\ell$ denotes the partition coefficient, defined as the ratio of the uniform bulk solute concentrations in the two media, $\ell=c_2^\infty/c_1^\infty$, with $c_1^\infty$ and $c_2^\infty$ denoting the respective background concentrations in media 1 and 2.
The first boundary condition accounts for the concentration discontinuity across the interface arising from the different solubilities of the chemical species in the two media. Accordingly, local equilibrium at the interface permits a finite concentration jump between the two phases, with its magnitude determined by $\ell$.
The second boundary condition expresses the continuity of the chemical flux normal to the interface, ensuring that there is no accumulation or depletion of solute at the interface. It follows from conservation of mass and requires the diffusive flux entering the interface from one medium to be equal to the flux leaving it into the other medium.
We further define the dimensionless ratio
\begin{equation}
    \lambda = \frac{D_2}{D_1} = \frac{\eta_1}{\eta_2} \, ,
\end{equation}
where the second equality follows from the Stokes–Einstein relation~\cite{einstein06}, assuming the same solute is considered in both media with comparable effective hydrodynamic size and temperature.
However, it is not generally valid for different solvents or solute–solvent interactions. In our formulation, $\lambda=D_2/D_1$ may therefore be regarded simply as the solute diffusivity ratio, with the Stokes–Einstein relation providing only a convenient physical interpretation rather than a necessary assumption.

The dynamics of phoretic active colloids can be obtained following a well-established theoretical framework~\cite{golestanian_les_houches}. In such systems, particles propel themselves through concentration gradients generated by their own surface activity. These gradients induce a tangential slip flow along the particle surface, given by~\cite{golestanian05,golestanian07},
\begin{equation}
    \bm{v}_\mathrm{S} = b 
    \left( \bm{I} - \hat{\bm{n}} \hat{\bm{n}} \right) \cdot \boldsymbol{\nabla} c \, ,
    \label{eq:vs_def}
\end{equation}
where $b$ denotes the local phoretic mobility and $\hat{\bm{n}}$ is the outward unit normal to the surface of the active colloid.

The induced phoretic velocity can be obtained without explicitly solving the surrounding hydrodynamic flow by invoking the reciprocal theorem of low-Reynolds-number fluid mechanics~\cite{masoud2019reciprocal}.
The key idea is to combine the original hydrodynamic problem with a suitably chosen auxiliary flow, whose known solution can then be used to determine the particle velocity. Here, we take the auxiliary problem to be that of a passive spherical particle subjected to a prescribed external force.
The procedure is classical and well established in the field~\cite{stone96, uspal15}, and we therefore refrain from providing the details here.

We characterize the far-field regime through the dimensionless ratio $\epsilon=R/d$, where $d=\rho\sin(\alpha-\beta)$. Our analysis assumes that the particle is small compared with its distance from the boundary, corresponding to $\epsilon\ll1$.
To leading order, the particle’s translational velocity in the auxiliary problem may be approximated by the corresponding result for an isolated particle in an unbounded fluid~\cite{ibrahim2016walls,daddi2022diffusiophoretic,daddi2026selfdiffusio}.
Specifically,
\begin{equation}
    \bm{V} = 
    -\frac{1}{4\pi R^2} \oint_\mathcal{S} \bm{v}_\mathrm{S} \, \mathrm{d}S \, ,
\end{equation}
with $\mathcal{S}$ denoting the surface of the isotropic active colloid.
We assume a uniform phoretic mobility over the particle surface. Under this assumption, the induced angular velocity vanishes, and the particle undergoes purely translational motion.

\subsection{Fourier-Kontorovich-Lebedev integral transform}

We employ the Fourier–Kontorovich–Lebedev transform, a powerful analytical technique for boundary-value problems posed in wedge geometries. Within this representation, the axial and radial coordinates are mapped onto the spectral variables $k$ and $\nu$, respectively, which greatly facilitates the solution procedure. While the Fourier transform is a familiar and widely used tool, the Kontorovich–Lebedev transform is comparatively less common. Originally introduced by Kontorovich and Lebedev in the context of boundary-value problems, it was subsequently developed and formalized in a number of classical references~\cite{lebedev1966problems, lebedev1972special, lebedev1979worked, erdelyi1953higher, yakubovich1996index}. Since then, it has become an established method for the analysis of wedge-domain problems across several disciplines, with applications spanning wave propagation~\cite{rawlins1999diffraction, antipov2002diffraction}, elasticity~\cite{daddi2025proc, daddi2025jelasticity}, and viscous hydrodynamics~\cite{sano1978effect, daddi2026hydrodynamic, daddi2026spectral}.

Throughout the remainder of this article, a tilde will be used to denote quantities expressed in the FKL transform space. The forward FKL transform is defined by
\begin{equation}
    \widetilde{f}(\nu,\theta,k) :=
    \mathscr{T}_{i\nu} \{f\} = 
    \int_{-\infty}^{\infty} \mathrm{d}z \, e^{ikz} \int_{0}^{\infty} f(r,\theta, z) \, \text{K}_{i\nu} ( |k|r) \, r^{-1} \, \mathrm{d} r \, .
\end{equation}
In this expression, $K_{i\nu}(|k|r)$ denotes the modified Bessel function of the second kind of imaginary order $i\nu$~\cite{abramowitz72}. The associated inverse transform is then expressed as
\begin{equation}
   f(r,\theta, z) = \mathscr{T}_{i\nu}^{-1} \{f\}
   = \frac{1}{\pi^3} 
   \int_{-\infty}^{\infty} \mathrm{d}k\, 
   e^{-ikz} 
   \int_0^\infty \widetilde{f}(\nu,\theta, k) \, \text{K}_{i\nu} \left(|k|r\right) \sh (\pi\nu) \, \nu \, \mathrm{d} \nu \, .
   \label{eq:inv-FKL-definition}
\end{equation} 
In a wide range of applications, the inverse Fourier transform can be evaluated analytically, reducing the solution to a single improper integral over the radial spectral variable $\nu$. We also note that, for $x>0$, the modified Bessel function $K_{i\nu}(x)$ is real-valued and possesses an even dependence on its order.

Upon applying the FKL transform, the Laplace equation for the concentration field reduces to a second-order ordinary differential equation in the polar angle $\theta$, of the form~\cite{daddi2025proc}
\begin{equation}
     \left( \frac{\partial^2}{\partial\theta^2}-\nu^2 \right) \widetilde{f} = 0 \, .
     \label{eq:Laplace_FKL_space}
\end{equation}
The general solution of Eq.~\eqref{eq:Laplace_FKL_space} can be written as
\begin{equation}
     \widetilde{f} 
     = 
     A \sh (\theta\nu) + A^\dagger \ch (\theta\nu) \, ,
     \label{eq:general_solution_form}
\end{equation}
where $A$ and $A^\dagger$ denote functions of $k$ and $\nu$ to be determined by enforcing the boundary conditions.

\section{Solute concentration field}

To construct the concentration field in the two media, we decompose each solution into an infinite-medium contribution and a set of correction terms introduced to account for the presence of the interface and confining boundaries. Specifically, we write
\begin{equation}
     c_1=c_1^\infty+C_0 + C_1 \, , \qquad c_2=c_2^\infty+C_2 \, . 
\end{equation}
The term $C_0$ denotes the free-space bulk concentration field generated by the source distribution in an unbounded medium, whereas $C_1$ and $C_2$ denote the correction fields introduced to enforce the boundary and interfacial conditions in media 1 and 2, respectively. This decomposition separates the known unbounded-medium solution from the additional contributions induced by the presence of interfaces and boundaries, allowing the remaining problem to be formulated in terms of the correction fields alone.

Using cylindrical coordinates, the free-space contribution takes the form
\begin{equation}
    C_0 = \frac{1}{s} \, , \qquad
    s =  \sqrt{r^2 + \rho^2 - 2\rho r \cos(\theta-\beta)+z^2} \, .
    \label{eq:c_free_space}
\end{equation}

The central idea of our approach is to first represent the known bulk solution given by Eq.~\eqref{eq:c_free_space} in FKL space, denoted by~$\widetilde{C}_0$, and then determine the complementary solutions required to satisfy the boundary conditions, also in FKL space, denoted by $\widetilde{C}_1$ and $\widetilde{C}_2$.
The corresponding solution in real space is subsequently recovered through the inverse transform.

We subsequently make use of the classical table of integral transforms compiled by Erdélyi \textit{et al.} \cite[p.~175]{erdelyi54} to evaluate the FKL transform of $1/s$, yielding
\begin{equation}
    \widetilde{C}_0 =
    \mathscr{T}_{i\nu} \left\{ \frac{1}{s} \right\}
    = \frac{2\pi}{\nu} \, \csh(\pi\nu) \, \ch\left( \left( \pi - |\theta - \beta| \right) \nu \right) \text{K}_{i\nu} (|k|\rho) \, .
    \label{eq:FKL-one-over-s}
\end{equation}
Employing the general solution form given by Eq.~\eqref{eq:general_solution_form}, the solution in the interior region is written as
\begin{equation}
    \widetilde{C}_1 = 
    \frac{2\pi}{\nu\sh(\pi\nu)}
    \left( \Lambda_1 \sh(\theta\nu) + \Lambda_1^\dagger \ch(\theta\nu) \right) \text{K}_{i\nu}(|k|\rho) \, .
\end{equation}
In exterior region, care must be taken to ensure that the solution satisfies the periodic boundary conditions $\widetilde{C}_2(\theta=\pi)=\widetilde{C}_2(\theta=-\pi)$ and $\partial_\theta\widetilde{C}_2(\theta=\pi)=\partial_\theta\widetilde{C}_2(\theta=-\pi)$, suggesting a solution of the form
\begin{equation}
    \widetilde{C}_2 = 
    \frac{2\pi}{\nu\sh(\pi\nu)}
    \left(
    \Lambda_2 
    \sh \left( \left( \pi-|\theta|\right)\nu\right) \operatorname{sgn}\theta
    + 
    \Lambda_2^\dagger 
    \ch \left( \left( \pi-|\theta|\right)\nu\right)
    \right) \text{K}_{i\nu}(|k|\rho) \, ,
\end{equation}
with sgn denoting the sign function.
Here, $\Lambda_j$ and $\Lambda_j^\dagger$, $j\in\{1,2\}$, depend only on the parameters $\nu$, $\alpha$, $\beta$, $\lambda$, and $\ell$.
We have extracted the common factor $1/\left( \nu\sh(\pi\nu)\right)$, which cancels upon evaluating the inverse FKL transform; cf.\ Eq.~\eqref{eq:inv-FKL-definition}.  
For convenience and to simplify the expressions for the coefficients, we introduce the following parameter 
\begin{equation}
    \Gamma = \frac{1-\lambda \ell}{1 + \lambda \ell} \, ,
\end{equation}
which we refer to as the interfacial parameter. For positive $\lambda$ and $\ell$, it satisfies $\Gamma \in [-1,1]$, with its sign and magnitude characterizing the contrast in the interfacial response between the two media.
This parameter provides a useful measure for characterising the interfacial contrast relevant to the dynamics of active particles near rigid, permeable, and fluid--fluid interfaces~\cite{malgaretti2018self,turk2025autophoretic}. In particular, $\Gamma$ is related to the interfacial parameter $\Lambda^c$ introduced in Ref.~\cite{turk2025autophoretic}, with $\Lambda^c$ corresponding to our $\Gamma$ and $\lambda_c$ to the diffusivity ratio $\lambda=D_2/D_1$. The parameter $\kappa_c$ in Ref.~\cite{turk2025autophoretic} is a solute permeability parameter, taking the values $0$ and $1$ for impermeable and permeable interfaces, respectively, and corresponds to our partition coefficient $\ell$, which can take any positive value in the present formulation.

Applying the boundary conditions at $\theta=\pm\alpha$ specified in Eq.~\eqref{eq:BCs} yields the following expressions for the coefficients
\begin{subequations}
    \label{eq:Lam_coeffs}
    \begin{align}
    \Lambda_1 &= \Gamma\Delta_+
    \sh(\beta\nu) \sh \left( 2(\pi-\alpha)\nu\right) \, , 
    \qquad\quad
    \Lambda_2 = \ell (\Gamma+1) \Delta_+ \sh(\beta\nu) \sh(\pi\nu) \, , \\
    \Lambda_1^\dagger &= \Gamma\Delta_-
    \ch(\beta\nu) \sh \left( 2(\pi-\alpha)\nu\right) \, ,  
    \qquad\quad
    \Lambda_2^\dagger = \ell (\Gamma+1) \Delta_- \ch(\beta\nu) \sh(\pi\nu) \, , 
    \end{align}
\end{subequations}
where we have defined
\begin{equation}
    \Delta_\pm^{-1} = \sh(\pi\nu) \pm\Gamma \sh \left( (\pi-2\alpha)\nu \right) \, .
\end{equation}

The unknown concentration fields follow from the inverse FKL transform defined in Eq.~\eqref{eq:inv-FKL-definition}. The resulting integral over $k$ can be evaluated in closed form using standard identities from classical integral tables, leaving a single integral with respect to $\nu$. We thus obtain
\begin{equation}
    C_1(r,\theta,z) = \frac{1}{\sqrt{\rho r}}
    \int_0^\infty 
    \left( \Lambda_1 \sh(\theta\nu) + \Lambda_1^\dagger \ch(\theta\nu) \right)
   \sch(\pi\nu) \, \mathrm{P}_{i\nu-\frac{1} {2}}(\ch \mu) \,\mathrm{d}\nu\, ,
   \label{eq:C1_int}
\end{equation}
for the interior region, and
\begin{equation}
   C_2(r,\theta,z) = \frac{1}{\sqrt{\rho r}}
    \int_0^\infty 
    \left( \Lambda_2 \sh((\pi-|\theta|)\nu) \sgn\theta + \Lambda_2^\dagger \ch((\pi-\theta)\nu) \right)
   \sch(\pi\nu) \, \mathrm{P}_{i\nu-\frac{1} {2}}(\ch\mu) \,\mathrm{d}\nu\, ,
    \label{eq:C2_int}
\end{equation}
for the exterior region, where $\mathrm{P}_n$ denotes the Legendre function of the first kind of degree $n$, evaluated at the argument $\ch\mu=(r^2+\rho^2+z^2)/(2\rho r)$, which ensures that $\mu\geq 0$. In deriving this expression, we have made use of the following standard integral formula
\begin{equation}
    \frac{4}{\pi^2} \int_0^\infty \mathrm{K}_{i\nu}(kr) \, \mathrm{K}_{i\nu}(k\rho)\cos(kz) \, \mathrm{d}k 
    =
     \frac{1}{\sqrt{\rho r}} \, 
     \sch(\pi\nu)
     \, \mathrm{P}_{i\nu-\frac{1}{2}} (\ch\mu)  \, .
     \label{eq:analytical_result_Kinu}
\end{equation}
This identity can be found, for example, in Prudnikov \textit{et al.}~\cite[p.~390, Eq.~2.16.36.2]{prudnikov1992integrals} or Gradshteyn and Ryzhik~\cite[p.~719, Eq.~6.672.3]{gradshteyn2014table}. 
The Legendre function can be represented by the following integral
\begin{equation}
    \mathrm{P}_{i\nu-\frac12} (\ch\mu) = 
    \frac{\sqrt{2}}{\pi} \,
    \operatorname{cth}(\pi \nu)
    \int_{\mu}^\infty 
\frac{\sin (\nu t) \, \mathrm{d}t}{\sqrt{\ch t-\ch\mu}} \, .
\label{eq:int_th_LegendreP}
\end{equation}
Accordingly, we obtain a double integral, which can be evaluated by first carrying out the integration with respect to $\nu$. Absolute convergence justifies the interchange of the order of integration. 
We cast the solution in the following form
\begin{equation}
    C_j = \frac{\sqrt{2}}{\pi \sqrt{\rho r}}
    \int_\mu^\infty \frac{\mathrm{d}t}{\sqrt{\ch t-\ch\mu}}\int_0^\infty f_j(\nu, t) \, \mathrm{d}\nu \, , \qquad j \in \{1,2\} \, .
    \label{eq:Cj_double_int}
\end{equation}
where the integrands $f_j$ can be obtained directly from the expression for $C_j$ given above.

To the best of our knowledge, an exact analytical evaluation of the resulting integrals with respect to the radial wavenumber~$\nu$ is not possible in the general case. Even when the semi-opening angle is commensurate, the resulting sums over the residues do not, in general, admit a closed-form analytical expression due to the presence of non-integer poles~\cite{scharstein2004green}. 
We therefore proceed by evaluating these integrals numerically.
This approach is robust in the present setting, since the integrands decay rapidly with increasing radial wavenumber. Nevertheless, exact analytical results can be obtained for commensurate opening angles in the limiting cases $\Gamma=\pm1$.

The improper integrals with respect to the radial wavenumber $\nu$ are evaluated numerically using the computer algebra system Mathematica and its \texttt{NIntegrate} routine~\cite{Mathematica}. \texttt{NIntegrate} employs an adaptive numerical integration algorithm that recursively subdivides the integration domain and concentrates numerical evaluations in regions where the estimated integration error is largest. In our calculations, the improper integrals are truncated at $\nu_{\max}=100$, beyond which the contribution is negligible. As demonstrated in our recent review article on a related problem involving the FKL transform~\cite{daddi2026spectral}, the truncation error decreases exponentially with the upper integration limit, and $\nu_{\max}=100$ generally yields an error of order $10^{-10}$. We have also verified that further increasing the truncation limit does not produce any appreciable change in the numerical results.

\begin{figure}
    \centering
    \includegraphics[width=\linewidth]{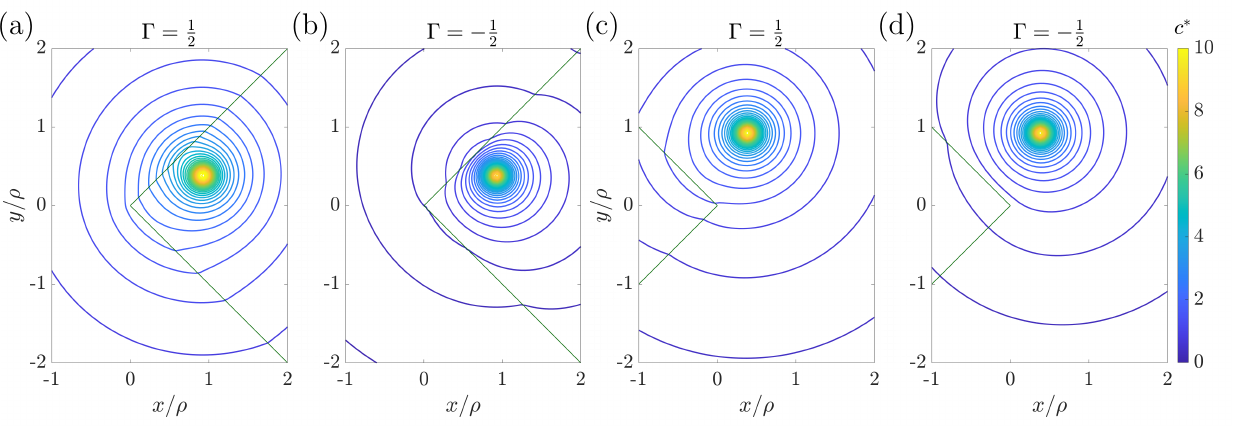}
    \caption{
    Contour plots of the scaled concentration field and corresponding iso-concentration lines generated by a source monopole in a two-medium wedge for $\alpha=\pi/4$ [(a) and~(b)] and $\alpha=3\pi/4$ [(c) and~(d)]. Results are shown for $\Gamma=1/2$ [(a) and~(c)] and $\Gamma=-1/2$ [(b) and~(d)]. The singularity is located at the polar angle $\beta=\alpha/2$, with partition coefficient $\ell=1$. 
    Results are shown in the plane $z/\rho=0.1$.
    All lengths are scaled by the radial distance $\rho$ of the singularity.
    }
    \label{fig:gam_pm_5}
\end{figure}

In Fig.~\ref{fig:gam_pm_5}, we present  contour plots of the scaled concentration field and corresponding iso-concentration lines generated by a source monopole in a two-medium wedge for wedge opening angles $\alpha\in\{\pi/4,3\pi/4\}$ and the interfacial parameter $\Gamma=\pm 1/2$.
Results are shown for $\ell=1$ in the plane $z/\rho=0.1$.
The concentration field exhibits a pronounced dependence on both $\alpha$ and $\Gamma$, which controls the transmission of angular concentration gradients through $\lambda=(1-\Gamma)/(1+\Gamma)$. In particular, $\Gamma=1/2$ corresponds to $\lambda=1/3$, resulting in a comparatively weaker angular variation in medium 1 and smoother, less distorted iso-concentration contours. In contrast, for $\Gamma=-1/2$, one has $\lambda=3$, resulting in a substantially stronger angular variation in medium 1 and a more pronounced redistribution of the concentration field across the interfaces. These interfacial effects are further modulated by the wedge geometry. 
Increasing the wedge opening angle reduces the geometric confinement and consequently leads to a weaker distortion of the concentration field, with the iso-concentration lines becoming progressively closer to the corresponding unbounded-fluid structure.
In both cases, the field retains the characteristic singular behavior near the source and decays progressively away from it, while the deformation of the iso-concentration lines provides a direct visualization of the coupling between wedge geometry and interfacial transport.

\subsection{Exact solution for $\Gamma=\pm1$ at commensurate wedge angles}

We now derive exact analytical expressions for the concentration field in medium 1 for the limiting cases $\Gamma=\pm1$ and commensurate wedge angles $\alpha$. The case $\Gamma=1$ corresponds to $\lambda\ell=0$, for which the interface imposes a no-flux boundary condition. Conversely, $\Gamma=-1$ is attained in the limit $\lambda\ell\to\infty$, in which the interfacial condition reduces to a Dirichlet condition, corresponding to a vanishing concentration perturbation at $\theta=\pm\alpha$.

For the interior domain, the corresponding integrands $f_\pm$ defined in Eq.~\eqref{eq:Cj_double_int} are given by
\begin{equation}
    f_\pm(\nu,t) = 
\bigl(
\sh\left(\left(\pi-2\alpha\right)\nu\right)
\ch\left(\left(\beta-\theta\right)\nu\right)
\csh\left(\pi\nu\right)
\pm \ch\left(\left(\beta+\theta\right)\nu\right)
\bigr)
\csh\left(2\alpha\nu\right)
\sin\left(\nu t\right) ,
\end{equation}
where the upper and lower signs correspond to $\Gamma=+1$ and $\Gamma=-1$, respectively.

We then express the integrands $f_\pm$ as linear combinations of terms whose corresponding integrals are available in standard integral tables.
The resulting integrals can subsequently be evaluated by applying the following closed-form formula
\begin{equation}
    \Psi(a,p,t) = \int_0^\infty\ch(a\nu)\csh(p\nu)  \sin(t\nu)\, \mathrm{d}\nu
    =
    \frac{\pi}{2p} \frac{ \sh \frac{\pi t}{p} }{\ch \frac{\pi t}{p} + \cos \frac{\pi a}{p} } \, .
    \label{eq:Psi_int_def}
\end{equation}
This integral can be found, for example, in the tables of Gradshteyn and Ryzhik~\cite[p.~510, Eqs.~3.981.8]{gradshteyn2014table}.
We obtain, after simplification and rearrangement of the terms,
\begin{equation}
\int_0^\infty f_\pm(\nu, t) \, \mathrm{d}\nu =
    \Psi \left( \theta_{-} + 2\alpha,2\alpha,t \right)
    \pm \Psi\left( \theta_+,2\alpha,t\right)
    -\Psi(\theta_{-} + \pi,\pi,t) \, ,
\end{equation}
where $\theta_\pm = \theta \pm \beta$.
The third term appearing in the latter expression, namely
\begin{equation}
    \Psi(\theta_{-}+\pi,\pi,t) = 
    \frac{1}{2} \frac{\sh t}{\ch t-\cos\theta_-} \, ,
\end{equation}
can be evaluated in closed analytical form when inserted into Eq.~\eqref{eq:C1_int}, whereas the remaining terms do not generally admit an analytical treatment. 
To obtain exact analytical results, we therefore restrict our attention to commensurate semi-opening angles, such that $\alpha=\pi/q$, where $q$ is a nonzero integer. In this case, the integration with respect to $t$ can be carried out exactly.
Depending on the parity of $q$, we seek an exact analytical expression for $\Psi$. 

\subsubsection{Even $q$: $q=2n$}

In this case, the analysis is considerably simpler owing to the symmetry of the setup.
We proceed by rewriting the remaining terms in the form
\begin{equation}
    \Psi(a,2\alpha,t) = 
    \frac{n}{2} \frac{ \sh (nt) }{ \ch (nt) + \cos (na) }
    = \frac{1}{2} \sum_{k=1}^q
    \frac{\sh t}{ \ch t - \cos\zeta_k } \, , 
    \label{eq:to_be_shown_appendix_1}
\end{equation}
where $\zeta_k = a+(2k-1)\pi/n$.
Technical details on the derivation of this series representation are provided in Appendix~\ref{appendix:chebyshev}.
Consequently, the solutions can be expressed on the inner domain as
\begin{equation}
     \int_0^\infty f_\pm(\nu, t)\, \mathrm{d}\nu = \frac{1}{2} \left(
    - \frac{\sh t}{\ch t-\cos\theta_-}
    \pm  \sum_{k=1}^n 
    \frac{\sh t}{\ch t -\cos \vartheta_k^+}
    + \sum_{k=1}^n 
    \frac{\sh t}{\ch t -\cos \vartheta_k^-} \right) ,
    \label{eq:Phi1_sum_even}
\end{equation}
where $\vartheta_k^+ = \theta_+ + (2k-1)\, \pi/n$
and $\vartheta_k^- = \theta_- + 2k\pi/n$.

We next substitute the integral given by Eq.~\eqref{eq:Phi1_sum_even} into Eq.~\eqref{eq:Cj_double_int} and evaluate the resulting improper integral with respect to $t$. The details of this calculation are provided in Appendix~\ref{appendix:int_1}, where we show that
\begin{equation}
    \int_{ \mu }^\infty 
    \frac{ \sh t }{\ch t - \cos \sigma} \frac{\mathrm{d}t}{\sqrt{ \ch t-\ch\mu }}
    =
    \frac{\pi}{ \sqrt{\ch\mu-\cos\sigma} } \, .
    \label{eq:int1}
\end{equation}

Finally, substituting these integral results into Eq.~\eqref{eq:Phi1_sum_even}, we obtain the following expressions for the solutions in the inner domain
\begin{equation}
    C_1 = \frac{1}{\sqrt{2\rho r}}
    \left( 
    \sum_{k=1}^n
    \frac{\pm 1}{\sqrt{\ch\mu-\cos\vartheta_k^+}}
    +
    \sum_{k=1}^{n-1}
    \frac{1}{\sqrt{\ch\mu-\cos\vartheta_k^-}}
    \right) ,
\end{equation}
where the first term outside the sums cancels with the last term of the second sum, thereby reducing the upper limit of the second sum to $n-1$.

\subsubsection{Odd $q$: $q=2n+1$}

The calculations become somewhat more involved when $q=2n+1$ is an odd integer; however, an exact result can still be obtained. We first express $\Psi$ as a finite series of the form
\begin{equation}
    \Psi(a,2\alpha,t) = 
    \frac{2n+1}{4} \frac{ \sh \left( \frac{2n+1}{2}\,t \right) }{ \ch \left( \frac{2n+1}{2}\, t \right) + \cos \left( \frac{2n+1}{2}\, a \right) }
    = \frac{1}{4} \sum_{k=1}^{2n+1}
    \frac{\sh \frac{t}{2}}{ \ch \frac{t}{2} - \cos\zeta_k } \, , 
\end{equation}
where $ \zeta_k = a/2 + \left( 2k-1\right)\pi/(2n+1)\, .$

Accordingly, the solutions can be written as
\begin{equation}
    \int_0^\infty f_\pm(\nu, t)\, \mathrm{d}\nu = 
    -\frac{1}{2} \frac{\sh t}{\ch t-\cos\theta_-}
    \pm\frac{1}{4} \sum_{k=1}^{2n+1}
    \frac{ \sh \frac{t}{2} }{ \ch \frac{t}{2} - \cos \vartheta_k^+ }
    +
    \frac{1}{4} \sum_{k=1}^{2n+1}
    \frac{\sh \frac{t}{2} }{ \ch \frac{t}{2} - \cos \vartheta_k^- }\, ,  
    \label{eq:Phi1_sum_odd}
\end{equation}
where we have defined $\vartheta_k^+ = \theta_+/2 + (2k-1) \, \pi/(2n+1)$ and $\vartheta_k^- = \theta_-/2 + 2k \, \pi/(2n+1)$.

It remains to substitute this expression into Eq.~\eqref{eq:Cj_double_int} and evaluate the resulting improper integral with respect to $t$. 
We show in Appendix~\ref{appendix:int_2} that
\begin{equation}
    \int_{ \mu }^\infty 
    \frac{ \sh \frac{t}{2} }{\ch \frac{t}{2} - \cos \sigma} \frac{\mathrm{d}t}{\sqrt{ \ch t-\ch\mu }}
    =
    \frac{2}{\sqrt{\ch\mu-\cos(2\sigma)}}
    \operatorname{acos}
    \left( 
    -\frac{\cos\sigma}{M}
    \right),
    \label{eq:int2}
\end{equation}
where $M= \ch (\mu/2)$.

Finally, upon substituting the integral results given by Eqs.~\eqref{eq:int1} and~\eqref{eq:int2} into Eq.~\eqref{eq:Cj_double_int}, the solutions for the inner domain take the form
\begin{equation}
    C_1 =
    \frac{1}{\pi \sqrt{ 2\rho r }}
    \left( 
    -\frac{\pi}{ \sqrt{\ch\mu-\cos\theta_-} }
    \pm \sum_{k=1}^{2n+1}
    \frac{\operatorname{acos}
    \left( 
    -\frac{\cos\vartheta_k^+}{M}
    \right)}{\sqrt{\ch\mu-\cos\left(2\vartheta_k^+\right)}}
    +
    \sum_{k=1}^{2n+1}
    \frac{ \operatorname{acos}
    \left( 
    -\frac{\cos\vartheta_k^-}{M}
    \right)}{\sqrt{\ch\mu-\cos\left(2\vartheta_k^-\right)}}
    \right).
    \label{eq:C1_final_ODD}
\end{equation}

It is worth noting that the sums could be rearranged so that they terminate at $n$, exploiting the underlying symmetry of the angular terms. However, we prefer to retain the current representation, with the summation extending up to $2n+1$, because it provides a more transparent connection with the original derivation and the underlying $2n+1$-fold angular discretization. This form also avoids introducing separate contributions arising from the symmetry reduction and keeps the expressions more compact and easier to interpret.

The concentration field admits an exact representation as a finite superposition of image singularities, with its structure depending fundamentally on the parity of the integer $q$ defining the wedge semi-opening angle $\alpha=\pi/q$. For even values, $q=2n$, the method of images closes after a finite number of reflections, leading to a particularly simple representation in terms of finite sums of Coulomb-type singularities located at the image positions. The image system consists of $n$ contributions associated with the $\vartheta_k^+$ family and $n-1$ contributions associated with the $\vartheta_k^-$ family, since one contribution from the latter coincides with the original monopole term and cancels identically. Consequently, the outer-domain solution involves only the $\vartheta_k^+$ family and retains a simpler structure.
In contrast, for odd values, $q=2n+1$, the image construction leads to a more delicate but still finite representation. Each image contribution is weighted by an additional angular factor involving $\operatorname{acos}(-\cos\vartheta_k^\pm/M)$, which arises from the incomplete closure of the reflection sequence associated with odd wedge geometries. The resulting solution contains $2n+1$ contributions from each image family, with each term incorporating the corresponding angular weighting factor.

Thus, although the analytical form of the concentration field differs according to whether $q$ is even or odd, both cases yield exact finite expressions. The parity of $q$ determines whether the reflection symmetry produces a purely algebraic image system or requires additional angular weighting factors, thereby controlling the complexity of the resulting Green's function representation.

\subsection{Planar-interface limit: $\alpha=\pi/2$}

Here, we recover the exact solution for the particular case of a planar boundary, corresponding to $\alpha=\pi/2$, for the general range $\Gamma\in[-1,1]$. In this case, the integrand functions $f_j$, $j\in \{1,2\}$, are given by
\begin{equation}
    f_1(\nu,t) = \Gamma \ch \left( \theta_+\nu\right) \csh\left( \pi\nu\right) \sin(\nu t) \, , \qquad
    f_2(\nu,t) = \ell (\Gamma+1)
    \ch \left( \left( \pi \operatorname{sgn}\theta-\theta_- \right)\nu \right) \csh(\pi\nu) \sin(\nu t) \, ,
\end{equation}
where, again, $\theta_\pm = \theta\pm\beta$.

By invoking the integral $\Psi$ defined in Eq.~\eqref{eq:Psi_int_def}, we readily obtain
\begin{equation}
    \int_0^\infty f_1(\nu,t)\, \mathrm{d}\nu 
    = \Gamma \, \Psi \left( \theta_+, \pi, t \right) 
    = \frac{\Gamma}{2} \frac{\sh t}{\ch t+\cos\theta_+}
\end{equation}
for the image solution, and 
\begin{equation}
    \int_0^\infty f_2(\nu,t)\, \mathrm{d}\nu 
    = \ell \left( \Gamma+1\right) \Psi \left( \pi\operatorname{sgn}\theta-\theta_-, \pi, t \right) 
    = \frac{\ell \left( \Gamma+1\right)}{2} \frac{\sh t}{\ch t-\cos\theta_-} 
\end{equation}
for the exterior domain. 
The final step consists of substituting these expressions into Eq.~\eqref{eq:Cj_double_int} and carrying out the integration with respect to $t$. Using the integral given in Eq.~\eqref{eq:int1}, we readily obtain
\begin{equation}
    C_1 = 
    \frac{\Gamma}{\sqrt{ 2\rho r \left( \ch\mu + \cos\theta_+ \right) }} = 
    \frac{\Gamma}{ \overline{s} }
\end{equation}
for the image solution, 
where
\begin{equation}
    \overline{s} = \sqrt{r^2 + \rho^2 + 2\rho r \cos(\theta+\beta)+z^2} \, .
\end{equation}
This expression corresponds to the Green's function of the image singularity with respect to the planar wall at $x=0$. The reflection maps the angular position of the singularity from $\beta$ to $\pi-\beta$, cf.~Eq.~\eqref{eq:c_free_space}, as reflected in the dependence of $\overline{s}$ on $\theta+\beta$.

For the exterior domain, the corresponding expression is given by
\begin{equation}
    C_2 = \frac{\ell (\Gamma+1)}{ \sqrt{2\rho r \left( \ch\mu-\cos\theta_- \right)} } = 
    \frac{\ell (\Gamma+1)}{s} \, .
\end{equation}
This result shows that, in the exterior domain, $C_2$ reduces to the free-space singularity~$C_0$, with the planar boundary affecting only its amplitude through the factor $\ell(\Gamma+1)$.

\subsection{Semi-infinite planar-interface limit: $\alpha=\pi$}

The planar interface separating two half-spaces corresponds to $\alpha=\pi$. For $\Gamma\neq\pm1$, the spectral coefficients vanish identically, yielding only the trivial solution; non-trivial solutions arise only for $\Gamma=\pm1$. Hence, the limits $\alpha\to\pi$ and $\Gamma\to\pm1$ do not commute.
In this case, the integrals of the functions $f_\pm$ can be expressed as
\begin{equation}
    f_\pm(\nu,t) = \pm \Psi\left( \theta_+,2\pi, t \right) - \Psi\left( \theta_-,2\pi,t \right) .
\end{equation}
Substituting the resulting expressions into Eq.~\eqref{eq:Cj_double_int} and using the integral identity in Eq.~\eqref{eq:int2}, we obtain
\begin{equation}
    C_1 = 
    \frac{1}{\pi \sqrt{2\rho r}}
    \left( 
    \pm\frac{ \operatorname{acos} \left( \frac{1}{M}\, \cos\frac{\theta_+}{2} \right) }{ \sqrt{\ch\mu-\cos\theta_+} }
    -
    \frac{ \operatorname{acos} \left( \frac{1}{M}\, \cos\frac{\theta_-}{2} \right) }{ \sqrt{\ch\mu-\cos\theta_-} } \right) .
\end{equation}
The same result follows from Eq.~\eqref{eq:C1_final_ODD} by setting $n=0$, in which case the sum over $k$ reduces to the single term $k=1$. Using $\operatorname{acos}(-x)=\pi-\operatorname{acos} x$, the term outside the sum cancels out.

\begin{figure}
    \centering
    \includegraphics[width=\linewidth]{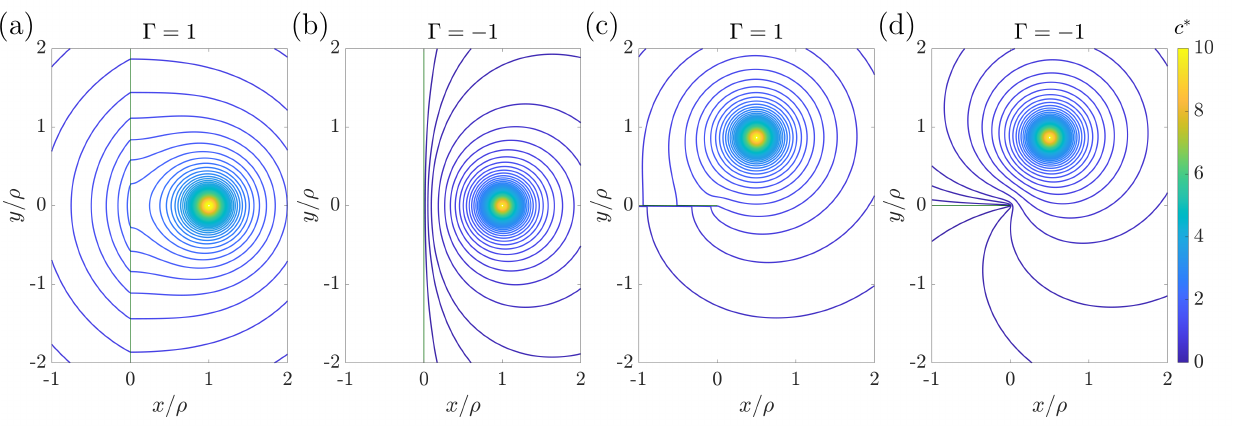}
    \caption{Contour plots of the scaled concentration field and corresponding iso-concentration lines generated by a source monopole near a planar wall at $x=0$, corresponding to $\alpha=\pi/2$, with the singularity located at $\beta=0$ [(a) and (b)], and near a semi-infinite planar wall along $y=0$ for $x\leq0$, corresponding to $\alpha=\pi$, with the singularity located at $\beta=\pi/4$ [(c) and (d)]. Results are shown for $\Gamma=1$ [(a) and (c)] and $\Gamma=-1$ [(b) and (d)]. 
    Results are shown in the plane $z/\rho=0.1$, with the partition coefficient set to $\ell=1$.
    All lengths are scaled by the radial distance $\rho$ of the singularity.
}
    \label{fig:planar}
\end{figure}

In Fig.~\ref{fig:planar}, we illustrate the limiting cases $\Gamma=\pm1$, corresponding to no-flux and zero-concentration boundary conditions, respectively, for planar and semi-infinite planar geometries.
Results are shown for $\ell=1$ in the plane $z/\rho=0.1$.
The concentration field exhibits markedly different behavior depending on the interfacial parameter $\Gamma$. For $\Gamma=1$, the concentration contours are smoothly reflected at the boundary, with the field remaining confined to the physical domain and exhibiting no normal concentration gradient at the interface. In contrast, $\Gamma=-1$ imposes a stronger constraint on the field. In particular, the concentration vanishes identically in the second domain, which is the region $x\leq0$ in panel (b), effectively reducing the problem to the corresponding single-domain Dirichlet problem. The same limiting behavior is reflected in the semi-infinite-wall geometry shown in panels (c) and (d), where the zero-concentration condition for $\Gamma=-1$ strongly suppresses the field near the boundary, whereas the no-flux condition for $\Gamma=1$ permits the concentration disturbance to extend smoothly up to the wall. Overall, the figure illustrates how the two limiting values of $\Gamma$ recover the familiar Neumann and Dirichlet boundary conditions while preserving the characteristic monopolar decay away from the singularity.

\section{Phoretic velocity}

Using the solution for the concentration field, the self-induced phoretic velocity can now be evaluated. The leading-order monopolar contribution is expected to scale as~$\epsilon^2$, where, again, $\epsilon=R/d$, with $d=\rho\sin(\alpha-\beta)$ denoting the perpendicular distance from the particle center to the upper interface.
Here, we focus on the case $\beta \in [0,\alpha)$ for simplicity, as the complementary case can be obtained by symmetry.

Enforcing the boundary condition at the interface, $\theta=\pm\alpha$, generally violates the boundary condition on the active colloid. A second reflection is therefore required to restore the latter. This reflection is singular at the particle position and contributes at the same leading order as the first. Specifically, the first reflection contributes two-thirds of the leading-order velocity, while the second contributes the remaining one-third. Omitting the second reflection thus gives the correct functional dependence but an overall prefactor smaller by a factor of $3/2$; see also Ref.~\onlinecite{malgaretti2018self}.

The partition follows directly from the surface-averaged tangential concentration gradient in Eq.~\eqref{eq:vs_def}. Denoting the contributions from the first and second reflections by $\bm{V}^{(1)}$ and $\bm{V}^{(2)}$, respectively, evaluation of the relevant harmonic on the particle surface gives
\begin{equation}
    \bm{V}^{(2)}=\frac{1}{2} \, \bm{V}^{(1)} \, .
\end{equation}
Hence, the leading-order velocity $\bm{V}=\bm{V}^{(1)}+\bm{V}^{(2)}$ satisfies
\begin{equation}
    \bm{V}^{(1)}=\frac{2}{3}\, \bm{V},
\qquad
\bm{V}^{(2)}=\frac{1}{3} \, \bm{V}\, .
\end{equation}
Thus, the $2/3$–$1/3$ partition results from the spherical concentration boundary condition and the structure of the first two solute reflections, with the second reflection providing the correction needed to satisfy that condition to leading order.

The phoretic slip velocity defined in Eq.~\eqref{eq:vs_def}, expressed in particle-centered spherical coordinates, is given by
\begin{equation}
    \bm{v}_\mathrm{S} = 
    \frac{b}{R} \left. \left( \frac{\partial c}{\partial \vartheta}\, \evarthe 
    + \csc\vartheta \, \frac{\partial c}{\partial \varphi} \, \evarphi
    \right) \right|_{s=R} ,
\end{equation}
which can be expressed in the Cartesian coordinate system as
\begin{equation}
    \bm{v}_\mathrm{S} = \frac{b}{R}
    \begin{pmatrix}
        \cos\vartheta \cos\varphi \, \cfrac{\partial c}{\partial \vartheta} - 
        \sin\varphi \csc\vartheta \,\cfrac{\partial c}{\partial \varphi} \\[3pt]
        \cos\vartheta \sin\varphi \, \cfrac{\partial c}{\partial \vartheta} + \cos\varphi \csc\vartheta \cfrac{\partial c}{\partial \varphi} \\[3pt]
        -\sin\vartheta \, \cfrac{\partial c}{\partial \vartheta}
    \end{pmatrix}_{s=R} .
    \label{vs_final}
\end{equation}

Due to the reflection symmetry of the wedge geometry about the $xy$ plane and the absence of any distinguished direction along the $z$-axis, the translational velocity is confined to the $xy$ plane. Accordingly, the particle velocity has no $z$-component, and can be expressed as
\begin{equation}
    \bm{V} = \frac{b}{4R}\, \epsilon^2 \, \sin^2 \left( \alpha-\beta \right)
    \int_0^\infty 
    \left( \Phi_x(\nu) \, \hat{\bm{e}}_x + \Phi_y(\nu) \, \hat{\bm{e}}_y \right) \mathrm{d}\nu \, , 
    \label{eq:V}
\end{equation}
where the integrands $\Phi_x$ and $\Phi_y$ are given by
\begin{subequations}
\begin{align}
\Phi_x &=
2\Gamma\Pi \left( \Delta_+ \sh(\nu\beta)
\left(
\cos(\beta)\sh(\nu\beta)
+2\nu\sin(\beta)\ch(\nu\beta)
\right)
+\Delta_-\ch(\nu\beta)
\left(
\cos(\beta)\ch(\nu\beta)
+2\nu\sin(\beta)\sh(\nu\beta)
\right) \right) , \\
\Phi_y &=
2\Gamma\Pi \left( \Delta_+ \sh(\nu\beta)
\left(
\sin(\beta)\sh(\nu\beta)
-2\nu\cos(\beta)\ch(\nu\beta)
\right)
+ \Delta_- \ch(\nu\beta)
\left(
\sin(\beta)\ch(\nu\beta)
-2\nu\cos(\beta)\sh(\nu\beta)
\right) \right) ,
\end{align}
\end{subequations}
where we have defined
\begin{equation}
    \Pi = \sh(2\alpha\nu)\sch(\pi\nu)
+2\sh\left(\nu(\pi-2\alpha)\right) .
\end{equation}

\subsection{Exact expressions for $\Gamma=\pm1$ at commensurate wedge angles}

An analytical evaluation of the integral is generally not possible in the general case because of the presence of non-integer poles. We therefore proceed numerically to compute the phoretic velocity. However, by exploiting the results derived above, analytical solutions can be obtained for $\Gamma=\pm1$ and for commensurate opening angles. In these cases, the integrals involving $\Phi_x$ and $\Phi_y$ are determined by the parity of $q$.

For $q=2n$, we obtain 
\begin{subequations} \label{eq:A_even}
    \begin{align}
    \int_0^\infty \Phi_x(\nu)\, \mathrm{d}\nu &= \pm
    \sum_{k=1}^n 
    \sin \phi_k
    \csc^2 \beta_k
    \kern0.1em +
    \sum_{k=1}^{n-1}
    \sin \left( \frac{k\pi}{n}-\beta \right)
    \csc^2 \frac{k\pi}{n} \, , \\
    \int_0^\infty \Phi_y(\nu)\, \mathrm{d}\nu &= \pm
    \sum_{k=1}^n 
    \cos \phi_k
    \csc^2 \beta_k
    +
    \sum_{k=1}^{n-1}
    \cos \left( \frac{k\pi}{n}-\beta \right)
    \csc^2 \frac{k\pi}{n} \, ,
\end{align}
\end{subequations}
where $\phi_k = (2k-1)\,\pi/(2n)$ and $\beta_k = \beta + \phi_k$.

For $q=2n+1$, the expressions take a more complex form, given by
\begin{subequations} \label{eq:A_odd}
    \begin{align}
    \int_0^\infty \Phi_x(\nu)\, \mathrm{d}\nu &=
    \pm\frac{1}{\pi}
    \left( 1 
    + \sum_{k=1}^n \left( X_k^+ + X_k^-\right)
    \right)
    + \cos\beta \left( \, \sum_{k=1}^n
    H_k - \frac{1}{\pi} \right), \\
    \int_0^\infty \Phi_y(\nu)\, \mathrm{d}\nu &= 
    \pm\frac{1}{\pi}
    \left(
    B -  \sum_{k=1}^n \left( Y_k^+ + Y_k^- \right)
    \right)
    \, + \, \sin\beta \left(\, 
    \sum_{k=1}^n
    H_k - \frac{1}{\pi} \right) ,
\end{align}
\end{subequations}
where we have defined the abbreviations
\begin{equation}
    B = \beta\csc^2\beta-\cot\beta \, , \qquad
    H_k = 2 \left( 1-\frac{2k}{2n+1} \right) \csc \frac{2k\pi}{2n+1} \, .
\end{equation}
In addition, 
\begin{subequations}
    \begin{align}
    X_k^\pm &= 
    \csc\beta_k^\pm
    \left( \sin\beta \mp \beta_k^\pm \csc\beta_k^\pm \sin  \phi_k \right) , \\[3pt]
     Y_k^\pm &= 
    \csc\beta_k^\pm
    \left(  \cos\beta+\beta_k^\pm \csc\beta_k^\pm \cos \phi_k \right) ,
\end{align}
\end{subequations}
where $\phi_k = (2k-1)\,\pi/(2n+1)$ and $\beta_k^\pm = \beta\pm\phi_k\mp \pi$.

For $\Gamma=\pm 1$, the expressions for the induced phoretic velocity exhibit distinct forms depending on the parity of $q$. For even $q=2n$, the integrals of $\Phi_x$ and $\Phi_y$ are expressed as finite sums involving trigonometric functions evaluated at equally spaced angular positions. In contrast, for odd $q=2n+1$, the expressions become more involved, containing additional contributions arising from the asymmetric angular distribution, together with auxiliary terms $X_k^\pm$, $Y_k^\pm$, and $H_k$. This distinction reflects the different geometrical symmetries associated with even and odd values of $q$.
In the limit $\beta\to0$, the transverse component of the induced phoretic velocity satisfies $V_y=0$ for both even and odd values of $q$, as required by the reflection symmetry of the system about the symmetry axis. In the odd case, this behavior is reflected by the fact that the additional contribution $B=\beta\csc^2\beta-\cot\beta$ appearing in the integral of $\Phi_y$ vanishes as $\beta\to0$.

\begin{table}[]
\centering
\renewcommand{\arraystretch}{2.75}
\setlength{\tabcolsep}{7pt}
\begin{tabular}{|c|c|c|c|c|c|c|}
\hline
$\alpha$ & 
$\dfrac{\pi}{8}$ &
$\dfrac{\pi}{6}$ &
$\dfrac{\pi}{4}$ & 
$\dfrac{\pi}{3}$ & 
$\dfrac{\pi}{2}$ & 
$\pi$ 
\\
\hline
$\displaystyle \left. V_x \middle/ \frac{b\epsilon^2}{4R} \right. $ & 
$\displaystyle \frac{1}{4}\left( 3\sqrt{2}-2 \right)\pm\sqrt{4-2\sqrt{2}} $ &
$\displaystyle \frac{\sqrt{3}}{3} \pm\frac{5}{4} $ &
$\displaystyle \frac{1}{2} \pm\sqrt{2}$ & 
$\displaystyle \frac{\sqrt{3}}{3}-\frac{3}{4\pi} \pm\left( \frac{3}{4\pi} + \frac{2\sqrt{3}}{3} \right)  $ &
$\pm 1$ & 
0 \\
\hline
\end{tabular}
\caption{
Scaled autophoretic translational velocity along the $x$-direction of a particle positioned on the midplane $(\beta=0)$ in a wedge geometry, evaluated analytically for various commensurate opening angles for $\Gamma=\pm1$.
}
\label{tab:V_beta0}
\end{table}

In Tab.~\ref{tab:V_beta0}, we provide the exact expressions for the induced phoretic velocity in the limiting cases $\Gamma=\pm1$, presented in scaled form for several wedge opening angles. The results are obtained by evaluating the finite series given by Eq.~\eqref{eq:A_even} for even values of~$q$, namely $q=2,4,6,$ and~8, and the series given by Eq.~\eqref{eq:A_odd} for odd values of~$q$, namely $q=1$ and~3. For these cases, the corresponding expressions simplify considerably. In particular, when $\Gamma=1$, which corresponds to imposing a no-flux (Neumann) boundary condition for the concentration on the wedge surfaces, the results reported in Tab.~1 of Ref.~\onlinecite{daddi2026selfdiffusio} are recovered exactly.

Hydrodynamic interactions between the two fluid media can contribute to the particle motion, potentially introducing a dependence of the phoretic velocity on the viscosity ratio. However, within the far-field expansion considered here, the leading hydrodynamic contribution to the translational velocity scales as $\sim\epsilon^3$ and is therefore weaker than the dominant phoretic contribution considered in the present analysis. For this reason, we have focused on the leading-order phoretic interaction and have not included the hydrodynamic correction.

The velocity expressions derived here are expected to be accurate when the separation is sufficiently large compared with the particle size. As the particle approaches the interface, higher-order reflections and finite-size corrections become increasingly important, and the present far-field expressions are no longer expected to provide a quantitatively accurate description. The far-field approximation is expected to remain reasonably accurate when the particle–interface separation is at least twice the particle radius, with its quantitative accuracy improving as the separation increases further. The near-interface regime, $d=O(R)$, lies outside the asymptotic regime considered here and would require a finite-size treatment or a higher-order reflection calculation.

\subsection{Planar-interface limit: $\alpha=\pi/2$}

In the limit $\alpha=\pi/2$, corresponding to a planar interface, the integrals of $\Phi_x$ and $\Phi_y$ simplify to
\begin{subequations}
    \begin{align}
    \int_0^\infty \Phi_x(\nu)\, \mathrm{d}\nu &= 2\Gamma \left( Z \cos\beta + \frac{\partial Z}{\partial \beta} \, \sin\beta\right) =\Gamma \sec^2\beta \, , \\
    \int_0^\infty \Phi_y(\nu)\, \mathrm{d}\nu &= 2\Gamma \left( Z \sin\beta - \frac{\partial Z}{\partial \beta} \, \cos\beta\right) = 0 \, ,
\end{align}
\end{subequations}
where we have made use of the fact that
\begin{equation}
    Z(\beta) = \int_0^\infty \ch(2\beta\nu) \sch(\pi\nu)\, \mathrm{d}\nu = 
    \frac{1}{2} \, \sec\beta \, .
\end{equation}
We thus recover known results obtained earlier with the velocity has a non vanishing component normal to the interface here along the $x$ direction~\cite{daddi2022diffusiophoretic}.
In particular, for $\Gamma=\pm1$, the first sum reduces to a single term corresponding to $k=n=1$, yielding $\pm\sec^2\beta$ for the integral of $\Phi_x$, while the corresponding integral of $\Phi_y$ vanishes.

\subsection{Semi-infinite planar-interface limit: $\alpha=\pi$}

The other interesting limit is the case $\alpha=\pi$, corresponding to a planar interface separating two half-spaces.
In this limit, the general solution becomes degenerate: for $\Gamma \neq \pm 1$, the spectral coefficients vanish identically, yielding a trivial solution. The non-trivial planar-interface solutions are recovered only for the limiting cases $\Gamma = \pm 1$. Thus, the limits $\alpha \to \pi$ and $\Gamma \to \pm 1$ are singular and cannot be interchanged.
In this limit, the integrals of $\Phi_x$ and $\Phi_y$ simplify to
\begin{subequations}
    \begin{align}
     \int_0^\infty \Phi_x(\nu)\, \mathrm{d}\nu &= \left( \pm U-U_0 \right)\cos\beta \pm \frac{\partial U}{\partial\beta}\, \sin\beta = \frac{1}{\pi} \left( \pm 1-\cos\beta \right) , \\
     \int_0^\infty \Phi_y(\nu)\, \mathrm{d}\nu &= \left( \pm U-U_0 \right)\sin\beta \mp \frac{\partial U}{\partial\beta}\, \cos\beta 
     = \frac{1}{\pi} 
     \left( \pm B - \sin\beta \right) ,
\end{align}
\end{subequations}
where we have defined
\begin{equation}
    U(\beta) = \int_0^\infty \ch(2\beta\nu) \sch^2(\pi\nu)\, \mathrm{d}\nu = 
    \frac{\beta}{\pi}\, \csc\beta \, , \qquad
    U_0 =\lim_{\beta\to 0}U(\beta) = \frac{1}{\pi} \, .
\end{equation}
An alternative route to obtaining these results is to use the expression given in Eqs.~\eqref{eq:A_odd} with $n=0$. In this case, the two sums make no contribution, and the velocities are determined solely by the terms outside the sums. This provides an independent check of the mathematical derivations presented for the case $\Gamma=\pm1$.

\subsection{Parallel-wall limit: $\alpha\to0$}

The limit $\alpha\to 0$ is particularly interesting, as it corresponds to the configuration of two parallel planar walls that are infinitely extended in the $(x,y)$ plane. We introduce the dimensionless parameter $\xi=\beta/\alpha\in[0,1]$, which characterizes the particle position across the channel. Specifically, $\xi=0$ corresponds to the mid-plane of the channel, while $\xi\to1$ corresponds to the upper wall. Denoting by $H$ the half-width of the channel and by $d$ the distance between the particle and the upper wall, we have
\begin{equation}
    \xi=1-\frac{d}{H}\, ,
\qquad
\epsilon=\frac{R}{d} \, ,
\end{equation}
where $\epsilon\ll1$.

In this limit, symmetry dictates that the phoretic velocity has no component along the $x$ direction, leaving only the wall-normal component, given by
\begin{equation}
    V_y = -\frac{2b\Gamma \epsilon^2}{R} \, 
    (1-\xi)^2
    \int_0^\infty \frac{\nu\sh(2\nu \xi)}
{e^{2\nu}-\Gamma^2e^{-2\nu}}\, \mathrm{d}\nu \, .
\end{equation}
Upon expanding the denominator as a geometric series in $\Gamma^2$, the integral can be evaluated term by term and expressed in terms of the Hurwitz–Lerch transcendent $\Phi$ as
\begin{equation}
    V_y = -\frac{b\Gamma \epsilon^2}{16R} \,
    (1-\xi)^2 
    \left( \Phi\left( \Gamma^2,2,\frac{1-\xi}{2} \right) - \Phi\left( \Gamma^2,2,\frac{1+\xi}{2} \right)  \right) .
\end{equation}
The corresponding infinite series is rapidly convergent and takes the form
\begin{equation}
    V_y = -\frac{b\epsilon^2}{R}\, \xi(1-\xi)^2 
    \sum_{n=0}^\infty 
    \frac{(2n+1) \Gamma^{2n+1}}{\left( (2n+1)^2-\xi^2 \right)^2} .
\end{equation}
Accordingly, the velocity vanishes at $\xi=0$, while its magnitude increases monotonically and approaches $-\Gamma/4$ as $\xi\to1$.
The latter limit corresponds to the single-wall limit, in which the influence of the second wall becomes negligible.

\begin{figure}
    \centering
    \includegraphics[width=\linewidth]{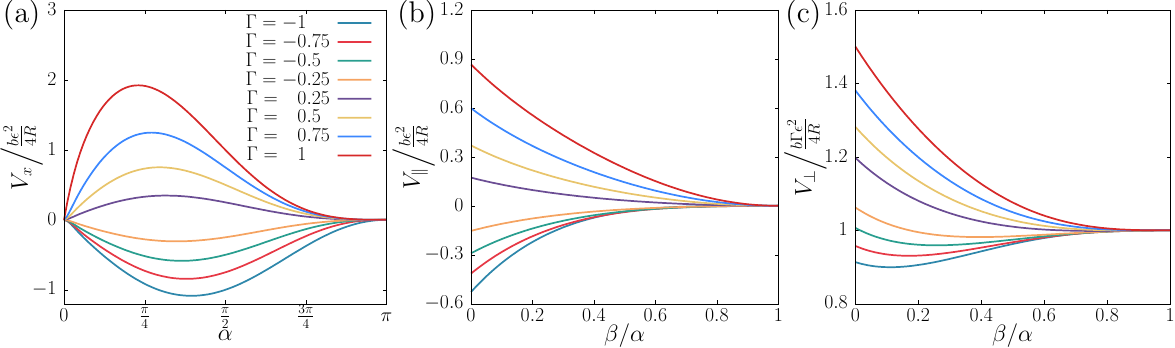}
    \caption{
Scaled translational velocity of the active particle as a function of its angular position within the wedge-shaped domain.(a)~Scaled velocity along the $x$-direction, as a function of the wedge semi-opening angle $\alpha$ for $\beta=0$ and eight values of the interfacial parameter $\Gamma\in[-1,1]$. (b)~Velocity component parallel to the upper interface, as a function of $\beta/\alpha$. 
(c)~Perpendicular velocity, rescaled by $\Gamma$, showing the approach to the single-interface limit as $\beta\to\alpha$. 
Here, $\alpha=\pi/3$ in panels~(b) and~(c), and the key in panel~(a) applies to all panels.
    }
    \label{fig:velo}
\end{figure}

Figure~\ref{fig:velo}~(a) shows the scaled translational velocity in the $x$-direction, for a particle located at $\beta=0$, as a function of the wedge semi-opening angle $\alpha$, for eight values of the interfacial parameter $\Gamma\in[-1,1]$. 
The velocity exhibits a non-monotonic dependence on the wedge geometry: it vanishes in both limiting configurations, $\alpha=0$ and $\alpha=\pi$, and attains a maximum magnitude at an intermediate value of $\alpha$. This behavior reflects a competition between the interfacial properties, encoded by~$\Gamma$, and the wedge geometry, which together determine the strength and distribution of the concentration gradients responsible for phoretic motion.  In the narrow-wedge limit $\alpha\to0$, the two boundaries approach one another, strongly constraining the concentration field and reducing the resulting phoretic motion. Conversely, as $\alpha\to\pi$, the wedge approaches the semi-infinite planar-boundary limit, in which the geometric contribution to the net translation along the $x$-direction vanishes. The strongest phoretic response therefore occurs at an intermediate opening angle, where the geometric asymmetry and confinement combine to generate the largest concentration gradients along the particle surface.

The sign of the velocity is controlled by the interfacial parameter $\Gamma$. For $\Gamma<0$, the particle translates in the negative $x$-direction, such that $V_x<0$, whereas for $\Gamma>0$, it moves in the positive $x$-direction, with $V_x>0$. Thus, reversing the sign of $\Gamma$ reverses the direction of propulsion. Moreover, the magnitude of $V_x$ increases systematically with increasing $|\Gamma|$, with the largest velocities occurring for $\Gamma=\pm1$. Physically, the parameter $\Gamma$ characterizes the contrast in the concentration response across the fluid--fluid interface. Increasing $|\Gamma|$ strengthens the influence of the interface on the concentration field generated by the particle, thereby enhancing the tangential concentration gradients and the associated phoretic slip. The figure consequently illustrates how the translational motion results from a competition between interfacial contrast and geometric confinement, with an optimal wedge opening at which the phoretic response is maximized.

Panels (b) and (c) of Fig.~\ref{fig:velo} show the components of the scaled translational velocity parallel and perpendicular to the upper interface of the wedge, respectively. These components are defined as $V_\parallel=V_x\cos\alpha+V_y\sin\alpha$ and $V_\perp=V_x\sin\alpha-V_y\cos\alpha$. The results are presented as functions of the normalized particle position $\beta/\alpha\in[0,1]$ for eight values of the interfacial parameter $\Gamma$, with the wedge semi-opening angle fixed at $\alpha=\pi/3$.

As shown in panel (b), the parallel velocity $V_\parallel$ varies monotonically with the particle position. Its magnitude is largest when the particle is located on the symmetry axis of the wedge, $\beta=0$, and decreases continuously as the particle approaches the upper interface, vanishing in the limit $\beta\to\alpha$. This behavior can be understood from the increasingly local character of the upper interface as the particle approaches it. In this limit, the particle effectively experiences a single planar boundary, which cannot generate a net translational motion parallel to itself for the present isotropic configuration. The parallel motion therefore originates from the combined influence of the two interfaces and is strongest when the particle is situated such that both sides of the wedge contribute appreciably to the concentration field. The sign of $V_\parallel$ is again dictated by the interfacial properties: $V_\parallel>0$ for $\Gamma>0$, whereas $V_\parallel<0$ for $\Gamma<0$. Thus, reversing the sign of $\Gamma$ reverses the direction of motion parallel to the interface, while increasing $|\Gamma|$ enhances its magnitude.

Panel (c) displays the perpendicular velocity $V_\perp$, rescaled by $\Gamma$. With this normalization, the different curves collapse onto the same limiting behavior as $\beta\to\alpha$, approaching unity. This limiting value corresponds to the particle approaching the upper interface, where the local geometry becomes equivalent to that of a single planar boundary, corresponding to the $\alpha=\pi/2$ limit. The collapse of the curves in this limit highlights the fact that the leading-order response near the interface is governed primarily by the local single-wall geometry, with the dependence on the interfacial parameter entering as an overall multiplicative factor. Consequently, the sign of $V_\perp$ is directly determined by the sign of $\Gamma$: $V_\perp>0$ for $\Gamma>0$ and $V_\perp<0$ for $\Gamma<0$. Together, panels (b) and (c) demonstrate that the two components of the particle velocity respond differently to the wedge geometry: the parallel component is intrinsically a two-interface effect and vanishes near a single interface, whereas the perpendicular component remains finite and approaches the corresponding single-wall behavior as the particle approaches the upper boundary.

\section{Conclusions}

This work has established an analytical framework for describing diffusiophoretic transport of a chemically isotropic active colloid in a three-dimensional wedge formed by two distinct fluid media. The concentration field is obtained by exploiting the Fourier--Kontorovich--Lebedev transform, which provides a natural representation for the coupled radial and axial dependence of the problem while allowing the interfacial conditions at the two wedge boundaries to be incorporated systematically. The resulting formulation provides an exact representation of the concentration field for arbitrary wedge opening angle and interfacial parameter $\Gamma$.

Particular analytical simplifications arise for the limiting cases $\Gamma=\pm1$ and for commensurate wedge angles. In these cases, the transform solution can be recast in terms of finite image constructions. The even and odd commensurability classes possess different structures, with the corresponding finite sums containing different numbers and arrangements of image contributions. These results provide explicit closed-form expressions for the concentration field and offer useful benchmarks for the general transform representation. For general values of $\Gamma$ and arbitrary wedge angles, the transform representation is evaluated numerically to determine the concentration field and the resulting phoretic velocity. The far-field expressions for the particle velocity are obtained asymptotically from the general solution, while the analytical results available for the special cases provide direct checks on the numerical evaluation. In addition, the general-$\Gamma$ solution recovers the known limiting configurations of a planar interface at $\alpha=\pi/2$ and a semi-infinite planar interface at $\alpha=\pi$, thereby providing further consistency checks on the formulation.

The concentration field was subsequently used to determine the leading-order translational phoretic velocity of the active colloid. The resulting motion is controlled by the interplay between the wedge geometry and the interfacial parameter $\Gamma$. For a particle located on the wedge symmetry axis, the velocity vanishes in the limiting geometries $\alpha\to0$ and $\alpha\to\pi$ and reaches its largest magnitude at an intermediate opening angle, demonstrating that geometric confinement can either enhance or suppress the phoretic response. The sign of $\Gamma$ determines the direction of motion, with opposite signs of $\Gamma$ producing opposite translational velocities, while increasing $|\Gamma|$ generally enhances the magnitude of the response. The velocity components parallel and perpendicular to the interface further reveal the distinct roles played by the two boundaries. In particular, the parallel component is a genuinely two-interface effect and vanishes as the particle approaches a single interface, whereas the perpendicular component approaches the corresponding single-interface behavior.
Exact expressions for the translational velocity were obtained for $\Gamma=\pm1$ and for several commensurate opening angles, as well as in the limiting geometries. In particular, the limit $\alpha\to0$ corresponds to two parallel interfaces and provides an additional useful benchmark for the general wedge solution. Together, these results demonstrate that the analytical formulation captures a broad range of geometries and interfacial conditions within a single framework.

To assess the relevance of the present theoretical framework to experimentally accessible phoretic systems, it is useful to consider representative ranges of particle size, velocity, and solute diffusivity.
Typical particle radii are $R\sim1$–$10\,\mu\mathrm{m}$, phoretic velocities are $V\sim 10^{-1}$–$10\,\mu\mathrm{m\,s^{-1}}$, and solute diffusivities are $D\sim10^{-9}$–$10^{-8}\,\mathrm{m^2\,s^{-1}}$. Assuming water at room temperature, with $\rho\sim10^3\,\mathrm{kg\,m^{-3}}$ and $\eta\sim10^{-3}\,\mathrm{Pa\,s}$, these values correspond to $\mathrm{Pe}=VR/D\sim10^{-5}$–$10^{-1}$ and $\mathrm{Re}=\rho VR/\eta\sim10^{-7}$–$10^{-4}$. Thus, fluid inertia is negligible throughout this parameter range, while the small-Péclet-number approximation is well justified provided $\mathrm{Pe}\ll1$. Regarding the geometric approximation, the present framework is applicable when the particle-interface separation is sufficiently large compared with the particle radius, i.e.\ $R/d\ll1$; in particular, particle-interface distances larger than twice the particle radius provide a regime in which the far-field approximation becomes increasingly appropriate.

The present analysis also suggests several directions for future work. An important extension would be to consider mixed boundary conditions in which one boundary is a fluid--fluid interface while the other is a no-flux boundary. Such a configuration is directly relevant to active particles confined within droplets pinned to substrates, where a fluid--fluid interface and a solid boundary meet to form a wedge-like environment near the contact region. Another natural extension would be to relax the assumption of an infinitely extended wedge and investigate finite-size effects. A finite wedge could introduce additional geometric length scales and modify both the concentration field and the resulting diffusiophoretic propulsion, potentially leading to confinement-induced effects that are absent in the present scale-free geometry.

More broadly, the results demonstrate that fluid--fluid interfaces and geometric confinement provide independent yet coupled means of controlling the propulsion of chemically isotropic active particles. The analytical solutions obtained here may therefore serve as useful benchmarks for numerical calculations and as a foundation for more realistic multiphase configurations encountered in experiments. The present analysis considers an idealized wedge of infinite axial extent, whereas practical microfluidic geometries have finite dimensions and may exhibit additional boundary effects, particularly near the channel ends and the wedge apex. Extensions incorporating finite axial extent, finite particle size relative to the confining geometry, nonuniform surface activity, more complex interfacial boundary conditions, and interfacial Marangoni stresses could therefore provide further insight into active-particle transport in confined multiphase environments. In particular, coupling phoretic solute transport to Marangoni flows may reveal additional mechanisms for modifying particle motion near fluid--fluid interfaces. Such developments may ultimately help connect idealized wedge geometries to experimentally realizable configurations and clarify how interfacial properties and confinement can be exploited to control the direction and magnitude of phoretic motion.

\begin{acknowledgments}
The author gratefully acknowledges R.\ Golestanian, A.\ M.\ Menzel, M.\ Lisicki, and S.\ Yakubovich for fruitful early collaborations on problems involving the Fourier--Kontorovich--Lebedev transform, which provided valuable foundations for the present work.
\end{acknowledgments}

\section*{Author contributions}
The author takes full responsibility for the content of this manuscript and has approved its submission.

\section*{Conflict of interest}
The author declares no conflict of interest.

\section*{Data Availability Statement}
The data supporting the findings of this study are available from the author upon reasonable request.

\section*{Funding}

No funding to acknowledge.

\appendix

\section{
Proof of the identity in Eq.~\texorpdfstring{\eqref{eq:to_be_shown_appendix_1}}{}
}
\label{appendix:chebyshev}

The series representation given by Eq.~\eqref{eq:to_be_shown_appendix_1} follows by exploiting the factorization properties of the Chebyshev polynomial of the first kind.
Using $T_n(\cos\theta)=\cos(n\theta)$ and $T_n(\ch t)=\ch(nt)$, we write
\begin{equation}
    T_n(x)+\cos(na)
=2^{n-1}\prod_{k=1}^{n}
\left(x-\cos\zeta_k\right) ,
\qquad
\zeta_k=a+(2k-1) \, \frac{\pi}{n} \, ,
\end{equation}
since $\cos(n\zeta_k)=-\cos(na)$. 
Setting $x=\ch t$ yields
\begin{equation}
    \ch(nt)+\cos(na)
=2^{n-1}\prod_{k=1}^{n}
\left(\ch t-\cos\zeta_k\right).
\end{equation}
Taking the logarithmic derivative with respect to $t$, we obtain
\begin{equation}
    \frac{n\sh(nt)}
{\ch(nt)+\cos(na)}
=
\sum_{k=1}^{n}
\frac{\sh t}{\ch t-\cos\zeta_k} \, .
\end{equation}

\section{Proof of the identity in Eq.~\texorpdfstring{\eqref{eq:int1}}{}}
\label{appendix:int_1}

The identity follows immediately from the substitution $u=\sqrt{\ch t-\ch\mu}$, so that $2u\,\mathrm du=\sh t\,\mathrm dt$.
Moreover, $t=\mu$ corresponds to $u=0$, while $t\to\infty$ gives $u\to\infty$. Hence
\begin{equation}
\int_{\mu}^{\infty}
\frac{\sh t}{\ch t-\cos\sigma}
\frac{\mathrm dt}{\sqrt{\ch t-\ch\mu}}
=
2\int_0^\infty
\frac{\mathrm du}
{u^2+\ch\mu-\cos\sigma}.
\end{equation}
Using
\begin{equation}
    \int_0^\infty\frac{\mathrm du}{u^2+a^2}
=\frac{\pi}{2a} \, ,
\qquad a>0 \, ,
\end{equation} 
with $a=\sqrt{\ch\mu-\cos\sigma}$
we obtain the desired integral. 

\section{Proof of the identity in Eq.~\texorpdfstring{\eqref{eq:int2}}{}}
\label{appendix:int_2}

Let $M=\ch(\mu/2)$ and $c=\cos\sigma$. Using $\ch t=2\ch^2(t/2)-1$ and the substitution $x=\ch(t/2)$, Eq.~\eqref{eq:int2} reduces to
\begin{equation}
    I=\sqrt{2}\int_M^\infty
\frac{\mathrm{d}x}{(x-c)\sqrt{x^2-M^2}}.
\end{equation}
Setting $x=M\ch u$ and subsequently $y=\operatorname{th}(u/2)$ gives
\begin{equation}
    I=2\sqrt{2}\int_0^1
\frac{\mathrm{d}y}{(M-c)+(M+c)\, y^2}
=
\frac{2\sqrt{2}}{\sqrt{M^2-c^2}}
\operatorname{atan}\sqrt{\frac{M+c}{M-c}} \,,
\end{equation}
where we have used the indefinite integral 
\begin{equation}
    \int \frac{\mathrm{d}y}{a+b y^2}
    =
\frac{1}{\sqrt{ab}}
\operatorname{atan}\left(\sqrt{\frac{b}{a}}\, y\right),
\qquad a,b>0 \, ,
\end{equation}
with $a=M-c$ and $b=M+c$.
Since $M>|c|$,
\begin{equation}
    2\operatorname{atan}\sqrt{\frac{M+c}{M-c}}
=\operatorname{acos}\left(-\frac{c}{M}\right),
\end{equation}
and therefore
\begin{equation}
    I=
\frac{\sqrt{2}}{\sqrt{M^2-c^2}}
\operatorname{acos}\left(-\frac{c}{M}\right).
\end{equation}
Finally, $\ch\mu-\cos(2\sigma)=2(M^2-c^2)$, yielding the desired result.

%


\end{document}